\documentclass[thmsa,12pt]{article}
\normalfont\sffamily
\usepackage{amsfonts}
\usepackage{amssymb}
\usepackage{amsmath}
\usepackage{amsthm}
\usepackage{graphicx}
\usepackage{textcomp}
\usepackage{a4wide}
\usepackage{chicago}
\usepackage{color}

\usepackage[normalem]{ulem}

\usepackage{pxfonts}

\usepackage{setspace}
\usepackage{footmisc}

\usepackage{hyperref}

\def\E{{\mathbf{E}}}

\renewcommand{\:}{\colon\!}
\def\l{\lambda}

\def\sigext{t}

\def\ordering{\varrho}

\newcommand{\realnumbers}{\mathbb{R}}

{\begin{list}{}{%
\setlength{\leftmargin}{0.5cm}%
\setlength{\rightmargin}{0.5cm}}%
\item[]\ignorespaces}%
{\unskip\end{list}}

\newtheoremstyle{mytheoremstyle}{}{}{\itshape}{
}{\scshape}{.}{\parindent}{}
\theoremstyle{mytheoremstyle}

\newtheorem*{theoremiid*}{Theorem \ref{thm:main_theorem}}
\newtheorem*{theoremshock*}{Theorem \ref{thm:main_result_for_shock_case}}
\newtheorem{proposition}{Proposition}
\newtheorem{corollary}{Corollary}
\newtheorem{lemma}{Lemma}
\newtheorem*{lemma*}{Lemma \ref{lem:convergence_of_shocked_densities}'}

\begin{document}

\def\title #1{\begin{center}
{\Large \sc #1}
\end{center}}

\def\author #1{\begin{center} {#1}
\end{center}}

\begin{titlepage}
\setstretch{1.3} \


\title{
Fair Representation and \\ a Linear Shapley Rule}

\author{Sascha Kurz\\ {\small Dept.\ of Mathematics, University of Bayreuth}}

\author{Nicola Maaser\\ {\small Economic Department, SOCIUM, University of Bremen\\ \mbox{[\emph{Corresponding Author}]}}}

\author{Stefan Napel\\ {\small Dept.\ of Economics, University of Bayreuth}
}

\begin{center} {\tt \today} \end{center}

\vspace{0.5cm}

\begin{center}{\bf {\sc Abstract}} \end{center}
{\small
When delegations to an assembly or council represent differently sized constituencies, they are often allocated voting weights which increase in population numbers (EU~Council, US~Electoral College, etc.).
The \emph{Penrose square root rule} (PSRR) is the main benchmark for `fair representation' of all bottom-tier voters in the top-tier decision making body, but rests on the restrictive assumption of independent binary decisions.
We consider intervals of alternatives with single-peaked preferences instead, and presume
positive correlation of local voters. This calls for a replacement of the PSRR by a \emph{linear Shapley rule}:
representation is fair if the Shapley value of the delegates is proportional to their constituency sizes.
}
\vspace{0.4cm}

\begin{description}
{\small
\item[Keywords:]
Shapley value; institutional design; two-tier voting; collective choice; 
equal representation; random order values
\item[JEL codes:]
    D02; 
    D63; 
    D70; 
    H77 
}
\end{description}
%

\setstretch{1.05}

\vfill
\noindent {\footnotesize
We are grateful for comments received at the European Public Choice Society Meeting 2016, Freiburg, and the 5$^\text{th}$~World Congress of the Game Theory Society, Maastricht.}

\end{titlepage}

\setstretch{1.3}

\section{Introduction}

\citeN{Shapley/Shubik:1954} advertised the {Shapley value} as ``A method for evaluating the distribution of power in a committee system'' almost immediately with the value's introduction by Lloyd~S. Shapley \citeyear{Shapley:1953}. Their motivation included not only the problem of measuring a~priori voting power in a given weighted voting system or in multicameral legislatures such as 
the US Congress, 
but they explicitly referred to the design of decision-making bodies and asked:
``Can a consistent criterion for `fair representation' 
be found?{''} (p.~787). This question was later taken up, and tentatively answered, by \citeN{Riker/Shapley:1968}.

Numerous studies in political science, economics, and business have since invoked the \emph{Shapley-Shubik index} (SSI)~-- which is simply the specialisation of the \emph{Shapley value} to the class of monotonic \emph{simple games} $(N,v)$ where $v\colon 2^N \to \{0,1\}$ categorizes coalitions $S\subseteq N$ as winning or losing according to a given decision rule. It has been used to examine the division of power in committees, shareholder meetings, councils, and assemblies or to assess the power shifts caused by EU enlargements, changes of treaties, etc. See \citeN{Felsenthal/Machover:1998}, \citeN{Laruelle/Valenciano:2008}, or \citeN{Holler/Nurmi:2013} for overviews.

This wide application in power analysis notwithstanding, the Shapley value's role as a tool for designing political institutions is probably outshone by a fairness benchmark which relates to the \emph{Banzhaf value}. The latter was introduced to the game theory community by \citeN{Dubey/Shapley:1979}, when they provided comprehensive mathematical analysis of a voting power index proposed by the lawyer John~F.\ Banzhaf \citeyear{Banzhaf:1965}. Banzhaf's interest in voting power was sparked by the US Supreme Court's series of `one person, one vote' rulings in the 1960s. His index was popularized in later legal cases. 

Without Shapley's, Shubik's, Riker's, and Banzhaf's knowing, an equivalent power measure had already been investigated, and the question of fair representation partly been settled by statistician Lionel~S. Penrose \citeyear{Penrose:1946,Penrose:1952}.\footnote{The measure was again independently proposed by social scientists \citeN{Rae:1969} and \citeN{Coleman:1971}. See \citeN{Felsenthal/Machover:2005} for the curious history of ``mis-reinvention'' in the analysis of voting power.}
With the newly established United Nations in mind, Penrose studied two-tier systems in which constituencies (members countries, states, etc.) of different sizes elect one delegate each to a decision-making assembly. He explained how proportional weighting would give voters in larger constituencies disproportionate power.
Rather, the problem of giving equal representation to all constituents is solved by choosing top-tier voting weights such that the resulting pivot probabilities (i.e., Banzhaf value) of the delegates is proportional to the square root of the represented population sizes. This result~-- re-derived by \citeN{Banzhaf:1965} and sketched by \citeN{Riker/Shapley:1968}~-- is now known as the \emph{Penrose square root rule} (PSRR). It and the corresponding \emph{Penrose-Banzhaf index} became the key reference for many applied studies on federal unions and two-tier voting systems such as the US~Electoral College (e.g., \citeNP{Grofman/Feld:2005}; \citeNP{Miller:2009}, \citeyearNP{Miller:2012}),
the Council of the EU (e.g., Felsenthal and Machover 
\citeyearNP{Felsenthal/Machover:2004};  \shortciteNP{Fidrmuc/Ginsburgh/Weber:2009}), or the IMF (e.g., \citeNP{Leech/Leech:2009}).

The PSRR follows straightforwardly from assuming that citizens at the bottom tier vote \emph{independently} of each other, with equal probabilities for a `yes' and a `no'. The objective is that each voter shall have the same probability to cast a decisive vote, i.e., to swing the local decision and thereby the global one.\footnote{For odd population size $n_i=2k+1$,  decisiveness of voter~$l$ inside constituency~$\mathcal{C}_i$ coincides with an even split between the $2k$ other voters. The probability of this event, with individual `yes' and `no'-decisions being equally likely and independent, is $2^{-2k}\cdot {2k \choose k}$. By Stirling's formula, this is approximately $\sqrt{2/(\pi\cdot n_i)}$. Decisiveness of constituency~$\mathcal{C}_i$ at the top tier is captured by its Penrose-Banzhaf index in this setup, which must hence be rendered proportional to $\sqrt{n_i}$ by choosing appropriate voting weights.}
It is often forgotten, however, that the rule lacks justification if voters' decisions are \emph{statistically dependent} (cf.\ \citeNP{Chamberlain/Rothschild:1981} or \shortciteNP{Gelman/Katz/Tuerlinckx:2002}) or for \emph{non-binary} decisions.

We here propose to replace Penrose's binary model of two-tier voting by a continuous median voter environment, also analyzed by \shortciteN{Kurz/Maaser/Napel:2016}. The model gives a simple explicit micro-foundation for using the SSI rather than the Banzhaf value in two-tier voting analysis. Moreover, one arrives at a \emph{linear Shapley rule} instead of Penrose's square root one.

Our voters are assumed to have single-peaked preferences over an \emph{interval} of alternatives, not merely `yes' and `no'. Their delegates represent the median preference of the constituency in the considered assembly.
This assembly applies a weighted majority rule.
One hence obtains policy outcomes which equal the \emph{ideal point of the median voter of the constituency whose delegate is the assembly's weighted median}.
If there are many voters and the ideal points of their preferences have a continuous distribution and \emph{positive correlation} within each constituency (while being independent across constituencies, as in Penrose's model), then the probability of a given delegate being decisive in the assembly asymptotically approaches the Shapley value of the delegate, not the Banzhaf value.
Because the influence of a given voter on the position adopted by his or her delegate is inversely proportional to the constituency population~-- not to the square root as for binary options~-- this implies:
if voters shall have the same indirect influence on outcomes, the weighted majority game among their delegates must have a \emph{Shapley value proportional to the represented population sizes}.

This linear Shapley rule does not require strong correlation of individual opinions provided constituencies are as sizable as in most real applications. The assumption of \emph{some} preference affiliation within the constituencies suffices. It seems at least as natural as that of statistical independence also from a constitutional a~priori perspective. In particular, if all voters were perfectly exchangeable then there should be no objection to redrawing the constituency boundaries. One could then design constituencies to be approximately equal in terms of population size and the question of which voting weights to use would become redundant.

The identified linear Shapley rule does generally not imply that voting weights be proportional to population sizes. This holds only in the limit as the number of constituencies increases (\citeNP{Neyman:1982}, \citeNP{Lindner/Machover:2004}). The analysis hence strictly refines the simple intuition that `one person, one vote' calls for weights themselves to be proportional to populations.\footnote{Several other quantities are possible bases for apportionment, too, besides total population. Alternatives discussed in the US Supreme Court's recent \emph{Evenwel} v. \emph{Abbott} ruling include the number of registered voters and the number of eligible voters.} One should identify voting weights such that the resulting majority game implies a desirable Shapley-Shubik index. That is, one needs to solve the \emph{inverse problem} of the SSI.

We will point to some more of the related literature on two-tier voting systems in the next section and then present our median voting model 
in Section~\ref{sec:Model}. We formalize the fair representation problem in Section~\ref{sec:fair_representation}. The main result is derived in Section~\ref{sec:PolarizedAnalysis} and we discuss practical aspects of it in  Section~\ref{sec:Discussion}. We conclude in Section~\ref{sec:Conclusion}.

\section{Related Literature}\label{sec:Literature}

Most research on the design of two-tier voting systems has maintained the basic \emph{binary framework} adopted by Penrose \citeyear{Penrose:1946,Penrose:1952}, \citeN{Banzhaf:1965}, and \citeN{Riker/Shapley:1968}. One major strand of literature~-- including \citeN{Chamberlain/Rothschild:1981}, \shortciteN{Gelman/Katz/Tuerlinckx:2002}, \shortciteN{Gelman/Katz/Bafumi:2004}, and \citeN{Kaniovski:2008}~-- has considered relaxations of the assumption that individual `yes' or `no' decisions are independent and uniformly distributed. It has turned out that Penrose's square root rule lacks robustness. In particular, strictly less concave weight allocations are necessary if decisions exhibit positive correlation within constituencies.

A second big strand has considered other objectives than fair representation. The most salient alternative is optimal representation in the sense of maximal \emph{utilitarian welfare}. \citeN{Barbera/Jackson:2006} and \citeN{Beisbart/Bovens:2007} derived that, generally speaking, constituencies' voting weights need to be proportional to the expected utilitarian importance they attach to an issue.
This means that welfare is maximized by square root weights in case of independent voter preferences but by proportional weights in case of perfect alignment within the constituencies.
\shortciteN{Koriyama/Laslier/Mace/Treibich:2013} relatedly considered welfare with the twist that a voter's utility is not additive across multiple issues but a strictly concave function of the frequency with which the collective `yes' or `no' decision conforms to the individually preferred outcome. This generally calls for voting weights to be strictly concave in constituency sizes.

Other studies have considered the \emph{majoritarian objective} of selecting two-tier voting weights which, in a suitable sense, bring the implied top-tier decisions as close as possible to the decisions which would have resulted in a single encompassing constituency, i.e., in a direct referendum. Clashes between the outcomes of direct and indirect democracy~--
instances of the so-called \emph{referendum paradox} (see, e.g., \citeNP{Nurmi:1998:SCW})~-- are impossible to avoid; a prominent case was the election of President Bush by the US~Electoral College against the popular majority in 2000.
\citeN{Felsenthal/Machover:1999} have investigated the `mean majority deficit' of a two-tier system, referring to the difference between the size of the popular majority camp and the number of citizens in favor of the assembly's decision. \citeN{Kirsch:2007} instead considered the mean quadratic deviation between the shares of `yes'-votes at the bottom and top tiers, while \shortciteN{Feix/Lepelley/Merlin/Rouet/Vidu:2008} sought to minimize the probability of the top-tier decision being at odds with the majority of citizens.
All three studies identified a key role for weight assignments that relate to the \emph{square root} of constituency sizes if voter opinions are \emph{independent} and \emph{identically distributed (i.i.d.)}. However,  \citeN{Kirsch:2007} and \shortciteN{Feix/Lepelley/Merlin/Rouet/Vidu:2008} give warning that \emph{correlated} opinions at the constituency level may call for proportionality to the numbers of represented voters.
This dichotomy was confirmed also in simulations by \citeN{Maaser/Napel:2012} which left the binary Penrose-Banzhaf framework. Their objective was to minimize expected distance between the positions of the decisive delegate at the top tier and the electorate's median voter in case of an \emph{interval} of policy alternatives.

For the same convex policy environment, which we will also study here, \citeN{Maaser/Napel:2007} and \shortciteN{Kurz/Maaser/Napel:2016} have turned to the original question of `fair representation'.
\footnote{Relatedly, \citeN{Maaser/Napel:2014} have extended the analysis of additive {utilitarian welfare} from binary to interval policy spaces in simulations. Cardinal details  matter somewhat, but the general pattern in the literature~-- square root vs.\ linear weighting rules for independent vs.\ correlated preferences~-- is confirmed once more.} If the ideal points which characterize voters' single-peaked preferences are i.i.d., fair weights become proportional to the square root of population sizes as the number of constituencies increases. In view of asymptotic results by \citeN{Lindner/Machover:2004}, this matches Penrose's 
original conclusion even though the respective square root findings obtain from the superposition of very different effects.
Crucially, voting weights proportional to constituency sizes quickly perform better if  positive preference correlation is introduced.

\citeN{Laruelle/Valenciano:2007} and \shortciteN{LeBreton/Montero/Zaporozhets:2012} also have considered two-tier systems for non-binary decisions. The latter investigated a setting where delegates vote on allocations of transferable utility and the \emph{nucleolus} of the corresponding TU~game captures the respective constituencies' shares of surplus.
The former have considered situations where the space of policy alternatives gives rise to a \emph{Nash bargaining problem} with an implicit unanimity presumption.
Their delegates need bargaining powers proportional to their constituencies' sizes in order to avoid biases.

\citeANP{Laruelle/Valenciano:2007} raise the possibility~-- but without an explicit model~-- that the delegate's bargaining power in the considered committee equals the Shapley value of the simple voting game induced by a given weight assignment. Then unbiased or `neutral' representation calls for weights such that the resulting Shapley value is proportional to the represented constituents, exactly as our Corollary~\ref{cor:equal_representation_shock_case} asserts below.

With this exception and that of \citeN{Riker/Shapley:1968}, the Shapley value or SSI has so far, to the best of our knowledge, not featured as a benchmark for fair two-tier voting systems~-- despite its frequent application in positive analysis of voting power. \citeANP{Riker/Shapley:1968} provided no explicit mathematical analysis in their article.
In the wake of the US~Supreme Court's decisions in \emph{Baker} v.~\emph{Carr} and \emph{Reynolds} v.~\emph{Sims}, they focused on the \emph{delegate model} of representation where each representative acts as a funnel for binary majority decisions in his or her constituency. They argued, but did not prove, that a square root rule based on the SSI solves the problem in this model.

Much more briefly, \citeN{Riker/Shapley:1968} also discussed the Burkean \emph{trustee model} of representation. In that, representatives are `free agents' who ``seek to satisfy the general interest'' (p.~211) of their constituency rather than the interests of the winning majority. Under the ad~hoc assumption that such a free agent's SSI can be divided among all his constituents in equal measure, Riker and Shapley concluded for this case that a representative's \emph{SSI needs to be proportional to the number of voters} in his or her constituency. Our analysis derives the same conclusion from an explicit delegate model. The key distinction to the setting of \citeN{Riker/Shapley:1968} is that we consider many rather than only two policy alternatives and  incorporate preference correlation at the constituency level.

\section{Two-Tier Median Voter Model}\label{sec:Model}

We assume the same median voter framework as \shortciteN{Kurz/Maaser/Napel:2016}, and partly draw on the presentation therein. Take a population of $n$ \emph{voters} and let $\mathfrak{C} = \{\mathcal{C}_1,\ldots,\mathcal{C}_m\}$ be a partition of it into $m<n$ \emph{constituencies} $\mathcal{C}_i$ with $n_i=|\mathcal{C}_i| > 0$ members each.
The preferences of each voter $l\in \{1, \ldots, n\}=\bigcup_i \mathcal{C}_i$ are assumed to be \emph{single-peaked} over a finite or infinite real interval $X\subseteq \realnumbers$, i.e., a  convex rather than binary {policy space}. The respective peaks or ideal points are taken to be  \emph{identically distributed} and mutually \emph{independent across constituencies}. However, we allow for a particular form of preference correlation within each constituency.

Specifically, the \emph{ideal point} $\nu^l$ of voter~$l$ in constituency $\mathcal{C}_i$ is conceived of as the realization of a continuous random variable
\begin{equation}\label{eq:nu_decomposed_with_t}
\nu^l=t\cdot \mu_i+\epsilon^l
\end{equation}
where $t\cdot \mu_i$ is a constituency-specific shock. Random variable $\mu_i$ has the same continuous distribution $H$ for any $i\in\{1,\ldots,m\}$, with a bounded density and finite variance $\sigma^2_H$. The scalar $t\ge 0$ parameterizes the similarity of opinions within the constituencies.
Voter-specific shocks $\epsilon^l$ account for individual political and economic idiosyncrasies. They are presumed to have the same continuous 
distribution $G$ for all $l\in \{1, \ldots, n\}$ with finite  variance $\sigma^2_G$. The respective density is assumed to be positive and continuous at $G$'s median. This rules out the possibility of a gap between `left' and `right' opinions, which would generate a binary model through the backdoor. Variables $\epsilon^1, \ldots, \epsilon^n$ and $\mu_1, \ldots, \mu_m$ are assumed mutually independent.

A given profile $(\nu^1, \ldots, \nu^n)$ of ideal points could reflect voter preferences in  abstract left--right spectrums or regarding specific one-dimensional variables such as the location or scale of a public good, 
an exemption threshold for regulation, a transfer level, etc.
Variance $\sigma^2_G$ is a measure of \emph{heterogeneity within each constituency}.
Variance $t^2\sigma^2_H$ of $t\cdot \mu_i$
is a measure of \emph{heterogeneity across constituencies}. Preferences in all constituencies vary between left--right, high tax--low tax, etc.\ in a similar manner, but the constituencies' ranges of opinion are typically located differently from an interim perspective. Still, all ideal points are a~priori \emph{distributed identically}, i.e., we adopt a constitutional `veil of ignorance' perspective which acknowledges that $\nu^l$ and $\nu^k$ are correlated with
coefficient $t^2\sigma^2_H/(t^2\sigma^2_H+\sigma^2_G)$ whenever $l,k \in \mathcal{C}_i$. 

On any given issue, a policy $x^*\in X$ is selected by an \emph{assembly of representatives} 
which consists of one delegate from each constituency.\footnote{The constituencies could equivalently have multiple delegates who cast a uniform \emph{bloc vote}, as in the US Electoral College.}
Without going into details regarding the procedure for within-constituency preference aggregation (bargaining, electoral competition, or a central mechanism) we assume that the preferences of $\mathcal{C}_i$'s representative coincide with the respective median preference of the constituency. So the location of the ideal point of representative~$i$ is
\begin{equation}\label{eq:l_i_is_C_is_median}
\l_i\equiv \textnormal{median\,}\{\nu^l\colon l\in \mathcal{C}_i\} = t\cdot \mu_i+\tilde \epsilon_i
\end{equation}
with
\begin{equation}\label{eq:tilde_epsilon_defined}
\tilde \epsilon_i = \text{median\,}\{\epsilon^l\colon l\in \mathcal{C}_i\}.
\end{equation}
We admittedly put aside at least two practical problems with this assumption. First, systematic abstention of certain social groups can drive a substantial wedge between the median voter's and the median citizen's preferences, and non-voters go unrepresented. Second, due to agency problems, a representative's position may differ significantly from his district's median.\footnote{\citeN{Gerber/Lewis:2004} provide empirical evidence on how district median preferences and partisan pressures jointly determine representatives' behavior. 
} 

In the assembly, constituency $\mathcal{C}_i$ has voting weight $w_i\ge 0$. Any coalition $S\subseteq \{1, \ldots, m\}$ of representatives which achieves a combined weight
$\sum_{j\in S} w_j$ above
\begin{equation}\label{eq:qm_definition_generalized}
\tilde q\equiv q \sum_{j=1}^m w_j \ \ \text{for\ \ $q\in [0.5,1)$ }
\end{equation}
is winning and can pass proposals to implement some policy $x\in X$. This voting rule is denoted by $[\tilde q;w_1, \ldots, w_m]$.

Now consider the random permutation of $\{1, \ldots, m\}$ that makes $\l_{k\: m}$ the $k$-th leftmost ideal point among the representatives for any realization of $\l_1, \ldots, \l_m$. That is,
$\l_{k\: m}$ is their $k$-th order statistic. We disregard the zero probability events of several constituencies having identical ideal points and define the random variable $P$ by
\begin{equation}\label{eq:P_defined}
P\equiv \min \Big\{j\in\{1,\ldots,m\}\colon \sum_{k=1}^j w_{k:m} > \tilde q\Big\}.
\end{equation}
Representative $P\:m$ will be referred to as the \emph{pivotal representative} of the assembly.

In the case of \emph{simple majority rule}, i.e., $q = 0.5$, the ideal point $\lambda_{P\:m}$ of representative $P\:m$ cannot be beaten by any alternative $x\in X$ in a pairwise vote, i.e., it is a so-called \emph{Condorcet winner} and in the \emph{core} of the voting game defined by ideal points $\l_1, \ldots, \l_m$, weights $w_1,\ldots, w_m$ and quota $\tilde q$.\footnote{Note that for $x^*$ determined in this way, no constituency's median voter has an incentive to choose a representative whose ideal point differs from her own one, that is, to misrepresent her preferences (cf.\ 
\citeNP{Nehring/Puppe:2007}).} We take
\begin{equation}\label{eq:outcome_equals_pivotal_point}
x^*\equiv\l_{P\:m}.
\end{equation}
to be the collective decision taken by the assembly. We do so also in the non-generic cases of the entire interval $[\lambda_{P-1\:m}, \lambda_{P\:m}]$ being majority-undominated in order to avoid inessential case distinctions.\footnote{A sufficient condition for the core to be single-valued under simple majority rule is that the vector of weights satisfies $\sum_{j\in S} w_j\neq  q^{m}$ for each  $S\subseteq \{1,\ldots,m\}$.
}

The situation under supermajority rules is somewhat less clear-cut. A relative quota $q > 0.5$ typically induces an entire interval of undominated polices, instead of a single Condorcet winner. Still, representative $P\: m$ defined by (\ref{eq:P_defined}) will be considered to be the assembly's decisive member. This can, e.g., be justified by supposing a Pareto inefficient legislative status quo $x^\circ\approx \infty$ and that formation of a winning coalition proceeds as in many motivations of the Shapley value:
it starts with the most enthusiastic supporter of change (member~$1\: m$ of the assembly), iteratively including more conservative representatives, and gives all bargaining power to the first~-- and least enthusiastic~-- member~$P\: m$ who brings about the required majority.
\footnote{Status quo $x^\circ$ might also vary randomly on $X$. Then 
the quantity $\pi_i(t)$ below captures $i$'s pivot probability conditional on policy change.
Justifications for attributing most or all influence in a committee to representative $P\: m$ in the supermajority case date back to \citeN{Black:1948:EA}.
The focus on the core's extreme points can be motivated, e.g., by distance-dependent costs of policy reform. } So equation~(\ref{eq:outcome_equals_pivotal_point}) generally identifies the policy outcome for the given quota.

\section{The Problem of Fair Representation} \label{sec:fair_representation}

The event $\{x^*=\nu^l\}$ of voter~$l$'s ideal point coinciding with the collective decision under these presumptions almost surely entails that small perturbations or idiosyncratic shifts of $\nu^l$ translate into identical shifts of $x^*$, i.e., $\partial x^*/\partial \nu^l=1$. Voter~$l$ can then meaningfully be said to \emph{influence}, be \emph{decisive} or \emph{critical} for, or even to \emph{determine} the collective decision. This event has probability
\begin{equation}
p^l\equiv\Pr(x^*=\nu^l),
\end{equation}
which depends on the joint distribution of $(\nu^1, \ldots, \nu^n)$ and the voting weights $w_1, \ldots, w_m$ that are assigned to the assembly members. Even though $p^l$ will be very small if the set of voters $\{1, \ldots, n\}$ is large, it would constitute a violation of the `one person, one vote' principle if $p^l/p^k$ differed substantially from unity for any $l,k\in\{1, \ldots, n\}$.

Our objective of achieving fair representation can thus be specified as follows. Given a partition $\mathfrak{C} = \{\mathcal{C}_1,\ldots,\mathcal{C}_m\}$ of $n$ voters into constituencies, and distributions $G$, $H$ and a parameter $t\ge 0$ which together describe heterogeneity of individual preferences within and across constituencies, we seek to find a  mapping from $n_1, \ldots, n_m$ to weights $w_1, \ldots, w_m$ such that each voter a~priori has an equal chance of determining the collective decision $x^*\in X$~-- that is, such that
\begin{equation}\label{eq:OPOV_condition}
    {p^l}/{p^k}\approx 1\, \textnormal{\ \emph{for all} } l, k\in \{1, \ldots, n\}.
\end{equation}

The model's statistical assumptions imply that $p^l=p^k$ holds for $l, k\in \mathcal{C}_i$ irrespective of which specific $G$, $H$, $t$, and voting weights $w_1, \ldots, w_m$ are considered. Namely, the continuity of $G$ entails that if $l\in \mathcal{C}_i$ then
\begin{equation}\label{eq:within_constituency_pivot_probabilities}
    \Pr(\nu^l=\l_i)=\frac{1}{n_i}.
\end{equation}
So an individual voter's probability to be his or her constituency's median and to determine $\l_i$ is \emph{inversely proportional} to constituency $\mathcal{C}_i$'s population size. This will need to be compensated via his or her delegate's voting power in the assembly.

The events $\{\nu^l=\l_i\}$ and $\{x^*=\l_i\}$ are independent. (The first one only entails information about the identity of $\mathcal{C}_i$'s median, not its location.) It follows that the probability $p^l$ for an individual voter~$l\in\mathcal{C}_i$ influencing the collective decision $x^*$ is $1/n_i$ times the probability of event $\{x^*=\l_i\}$ or, equivalently, of $\{P\:m = i\}$.
Letting
\begin{equation}
\pi_i(t)\equiv \Pr(P\:m = i)
\end{equation}
denote the probability of constituency $\mathcal{C}_i$'s representative being pivotal in the assembly for a given parameter~$t$, a solution to the \emph{problem of fair representation} hence consists in mapping constituency sizes $n_1, \ldots, n_m$ to voting weights $w_1, \ldots, w_m$ such that
\begin{equation}\label{eq:OPOV_condition_restated}
    \frac{\pi_i(t)}{\pi_j(t)}\approx \frac{n_i}{n_j} \textnormal{ \emph{ for all} } i, j\in \{1, \ldots, m\}.
\end{equation}

Note that \emph{if} the representatives' ideal points $\l_1, \ldots, \l_m$ were not only mutually independent but also had identical distributions $F_i=F_j$ for all $i,j\in \{1, \ldots, m\}$ then all orderings of $\l_1, \ldots, \l_m$ would be equally likely.
In this situation, player~$i$'s probability of being pivotal $\pi_i(t)$ would simply be $i$'s \emph{Shapley value} $\phi_i(v)$ (see \citeNP{Shapley:1953}), where $v$ is the characteristic function of the $m$-player TU~game in which the worth $v(S)$ of a coalition $S\subseteq \{1, \ldots, m\}$ is 1 if $\sum_{j\in S} w_j>\tilde q$ and 0 otherwise,
and
\begin{equation}\label{eq:Shapley}
\phi_i(v)\equiv \sum_{S\subseteq \{1, \ldots, m\}}
 \frac{|S|!\cdot(m-|S|-1)!}{m!}[v(S\cup \{i\})-v(S)].
\end{equation}
Yet, under the normatively attractive `veil of ignorance' assumption that \emph{individual voters' ideal points} are identically distributed, the ideal points $\lambda_1,\ldots, \lambda_m$ of their representatives will only have an identical distribution in the trivial case $n_1=\ldots=n_m$. Otherwise, a smaller number $n_i<n_j$ of draws from the same distribution generates a sample whose median $\l_i$ has greater variance than the respective sample median $\l_j$.
Technically, $\pi(t)$ corresponds to a \emph{random order value} where the `arrival time' distributions are mean-preserving spreads of each other (see, e.g., \citeNP[Sec.~4]{Monderer/Samet:2002}).

\section{Fair Representation with Affiliated Constituencies}\label{sec:PolarizedAnalysis}

\shortciteN{Kurz/Maaser/Napel:2016}, for a simple majority quota $q=0.5$ in the assembly, study how the sample size effect on the realized medians gives a pivotality advantage to the delegates from large constituencies. For instance, in the i.i.d.\ case with $t=0$, $n_j=4\cdot n_i$ implies that the delegate from constituency $\mathcal{C}_j$ is twice as likely to be the unweighted median among the delegates, i.e., $\pi_j(0)=2\cdot \pi_i(0)$, if $n_i$ is sufficiently big. A fair allocation then needs to give delegate~$j$ only about twice the weight of delegate~$i$ in order to satisfy (\ref{eq:OPOV_condition_restated}). More generally, the observation that the density of the sample median $\l_i$ at the expected location of $\l_{P:m}$ is proportional to the square root of sample size $n_i$ gives rise to a \emph{square root rule} as $m\to \infty$ in case $t=0$. See \shortciteN[Sec.~4]{Kurz/Maaser/Napel:2016} for details.

We here study the case $t>0$ and keep the number $m$ of constituencies fixed. We thus capture the realistic scenario in which a big electorate is partitioned into moderately many constituencies. These differ not just in size but exhibit some internal similarity.

The key observation for this case of internally affiliated constituencies is that the indicated sample size effect for the distribution of the median voter only pertains to the idiosyncratic components of delegates' preferences, i.e., $\tilde \epsilon_i = \text{median\,}\{\epsilon^l\colon l\in \mathcal{C}_i\}$. In particular, $\tilde \epsilon_i$'s variance is approximately $\tfrac{1}{2}\pi\cdot \sigma_G^2/n_i$  (see, e.g., \shortciteNP[Thm.~8.5.1]{Arnold/Balakrishnan/Nagaraja:1992}). This contrasts with a constant variance of $t^2\sigma_H^2$ for the constituency-specific preference component.

The density function of delegate~$i$'s ideal point $\l_i$ is the convolution of densities of a random variable that does not vary in $n_i$ and a random variable that vanishes in $n_i$. On the realistic presumption that $\sigma_G^2$ is not bigger than $\sigma_H^2$ by several orders of magnitude, the distribution of the constituency-specific shocks hence comes to dominate the distribution of individual-specific shocks as we consider population sizes in the thousands or millions.

Since our model conceives of the population distribution $n_1, \ldots, n_m$ as fixed, we will not consider limits as $n_i\to \infty$ for $i\in \{1,\ldots, m\}$. The phenomenon of $t\cdot \mu_i$'s variation dominating that of $ \tilde \epsilon_i$ for all $i\in \{1, \ldots, m\}$ is captured equally well by letting $t$ grow for given (population-dependent) variances of $\tilde \epsilon_1, \ldots, \tilde \epsilon_m$. We have the following formal result:

\begin{proposition}\label{thm:main_result_for_shock_case}
Consider an assembly with $m$ constituencies and relative decision quota $q\in [0.5; 1)$. Let the ideal point of each representative $i\in \{1, \ldots, m\}$ be $\l_i=t\cdot \mu_i+\tilde \epsilon_i$, and suppose
$\mu_1,\ldots, \mu_m$ and $\tilde \epsilon_1,\ldots,\tilde \epsilon_m$ are mutually independent,
$\tilde \epsilon_1,\ldots,\tilde \epsilon_m$ have finite second moments, and
$\mu_1,\ldots, \mu_m$ have identical bounded densities.
Then
\begin{equation}\label{eq:shockthm}
\lim_{\sigext\to \infty}\frac{\pi_i(t)}{\pi_j(t)} = \frac{\phi_i(v)}{\phi_j(v)}
\end{equation}
where $\phi(v)$ denotes the Shapley value of weighted voting game $v=[\tilde q; w_1, \ldots, w_m]$ and we suppose $\phi_j(v)>0$.
\end{proposition}

\begin{proof}
The proposition follows from the Shapley value's definition (\ref{eq:Shapley}) and the observation that the orderings which are induced by realizations of vectors $\boldsymbol{\l}=(\l_1, \ldots, \l_m)$ and $\boldsymbol{\mu}=(\mu_1, \ldots, \mu_m)$
will coincide with a probability that tends to $1$ as $t$ grows.
To see the latter, ignore any null events in which several shocks or ideal points coincide and let $\hat\ordering(\mathbf{x})$ denote the permutation
of $\{1,\dots,m\}$ such that $x_i<x_j$ whenever $\hat\ordering(i)<\hat\ordering(j)$ for a real vector $\mathbf{x}=(x_1,\ldots, x_m)$. We then have:

\begin{lemma}
Let $\l_i^t\equiv t\cdot \mu_i+\tilde \epsilon_i$ for $i\in\{1, \ldots, m\}$ where
$\mu_1,\ldots, \mu_m$ and $\tilde \epsilon_1,\ldots,\tilde \epsilon_m$ are mutually independent,
$\tilde \epsilon_1,\ldots,\tilde \epsilon_m$ have finite second moments, and
$\mu_1,\ldots, \mu_m$ have identical bounded densities.
Then
\begin{equation}
  \lim\limits_{t\to\infty} \Pr(\hat\ordering(\boldsymbol{\lambda}^t)=\ordering)=\lim\limits_{t\to\infty} \Pr(\hat\ordering(\boldsymbol{\mu})=\ordering)=\frac{1}{m!}
\end{equation}
for each permutation $\ordering$ of $\{1,\dots,m\}$.
\end{lemma}

\noindent To prove the lemma, denote the finite variance of $\tilde \epsilon_i$ by $\sigma_i^2$ and let
$U\equiv \left(\max_i |\E[\tilde\epsilon_i]|\right)^3$.
We can choose a real number $k$ such that the bounded density function $h$ of $\mu_i$, with $i\in \{1, \ldots, m\}$,
satisfies $h(x)\le k$ for all $x\in\mathbb{R}$. For any given realization $\mu_j=x$, the probability of the independent
random variable $\mu_i$ assuming a value inside interval $(x-4t^{-\frac{2}{3}}, x+4t^{-\frac{2}{3}})$
is bounded above by $k\cdot 8 t^{-\frac{2}{3}}$. We can infer that the event
$\big\{|\mu_i-\mu_j|<4 t^{-\frac{2}{3}}\big\}$, which is equivalent to the event
$\big\{|t\mu_i-t\mu_j|<4 t^{\frac{1}{3}}\big\}$, has a probability of at most $k\cdot 8 t^{-\frac{2}{3}}$
for any $i\neq j\in \{1, \ldots, m\}$. And we can conclude from Chebyshev's inequality that
$\Pr(|\tilde\epsilon_i-\E[\tilde\epsilon_i]|< t^{\frac{1}{3}})$ is at least
$1-\sigma_i^2\cdot t^{-\frac{2}{3}}$.
For $t\ge U$, we have $|\E[\tilde\epsilon_i]|\le t^{\frac{1}{3}}$; and if $|\tilde\epsilon_i-\E[\tilde\epsilon_i]|< t^{\frac{1}{3}}$ holds then also
\begin{equation}\label{ie:absolute_of_epsilon_bound}
2t^{\frac{1}{3}}>|\E[\tilde\epsilon_i]| + |\tilde\epsilon_i-\E[\tilde\epsilon_i]| \ge |\tilde\epsilon_i|
\end{equation}
by the triangle inequality. Hence, the probability for (\ref{ie:absolute_of_epsilon_bound}) to hold when $t\ge U$ is $\Pr(|\tilde\epsilon_i|< 2t^{\frac{1}{3}})\ge 1-\sigma_i^2\cdot t^{-\frac{2}{3}}$ for each $i\in \{1, \ldots, m\}$.

Now consider the joint event that (i) $|t\mu_i-t\mu_j|\ge 4t^{\frac{1}{3}}$ for \emph{all} pairs
$i\neq j\in \{1, \ldots, m\}$ and (ii) that $|\tilde\epsilon_i|< 2t^{\frac{1}{3}}$ for \emph{all}
$i\in \{1, \ldots, m\}$. In this event, the ordering of $\l_1^t, \ldots, \l_m^t$ is determined entirely by the
realization of $t \mu_1, \ldots, t \mu_m$; in particular, $\hat\ordering(\boldsymbol{\l}^t)=\hat\ordering(\boldsymbol{\mu})$.
Using the mutual independence of the considered random variables this joint event must have a probability of at least
\begin{equation}
  \prod\limits_{s=1}^{m \choose 2} \left(1-k\cdot 8 t^{-\frac{2}{3}}\right) \cdot \prod\limits_{i=1}^m
  \left(1-\sigma_i^2\cdot t^{-\frac{2}{3}}\right)
  \ge 1- \left(8 k{m \choose 2}+\sum_{i=1}^m \sigma_i^2\right)\cdot t^{-\frac{2}{3}}
\end{equation}
for $t\ge U$.
The right hand side tends to $1$ as $t$ approaches infinity. It hence remains to acknowledge that any
ordering $\hat\ordering(\boldsymbol{\mu})$ has an equal probability of $1/m!$ because $\mu_1, \ldots, \mu_m$ are i.i.d.
\end{proof}

We remark that Proposition~\ref{thm:main_result_for_shock_case} does \emph{not} require identity~(\ref{eq:l_i_is_C_is_median}) to hold; the limit (\ref{eq:shockthm}) applies also if $\l_i$ is determined, e.g., by an oligarchy instead of the median voter of $\mathcal{C}_i$.
Moreover, it is worth noting that Proposition~\ref{thm:main_result_for_shock_case} imposes very mild conditions on densities
$g_1, \ldots, g_m$, voting weights $w_1, \ldots, w_m$, or quota~$\tilde q$:
the Shapley value $\phi(v)$ automatically takes care of the combinatorial particularities associated with $[\tilde q; w_1, \ldots, w_m]$; and the convolution with $t\cdot \mu_i$'s bounded density, $\frac{1}{t} h\left(\frac{x}{t}\right)$, is sufficient to `regularize' any even non-continuous distribution $G_i$ of $\tilde \epsilon_i$.

Of course, applied to our two-tier median voter model, variables $\l_1,\ldots,\l_m$ are defined by (\ref{eq:l_i_is_C_is_median}) and $\tilde \epsilon_1,\ldots,\tilde \epsilon_m$
correspond to the medians of $n_1, \ldots, n_m$ draws of i.i.d.\ idiosyncratic preference components $\epsilon^l$.
The proposition then implies:

\begin{corollary}[Linear Shapley rule]\label{cor:equal_representation_shock_case}
If individual ideal points are the sum of i.i.d.\ idiosyncratic components 
and i.i.d.\ constituency components 
with similar orders of magnitude
then
\begin{equation}
(w_1, \ldots, w_m) \textnormal{\ \,such that\ \,} \phi(\tilde q;w_1,\ldots, w_m) \propto \left({n_1},\ldots, {n_m}\right)
\label{eq:shock_case_solution}
\end{equation}
achieves approximately fair representation for any given relative decision quota $q\in [0.5;1)$ if
constituency populations are large.
\end{corollary}

\section{Discussion}\label{sec:Discussion}

Given a particular weighted voting scheme, various techniques can be used to compute the Shapley value or voters' SSI efficiently. Implementation of the linear Shapley rule, however, requires solving the more challenging \emph{inverse problem}: find a weighted voting game that (approximately) induces a desired SSI vector. Specifically, given a relative threshold $q$ and denoting relative population sizes by $\mathbf{\bar n} = (n_1, \dotsc, n_m)/n$, the linear Shapley rule requires finding a solution to the minimization problem
\begin{equation}\label{eq:inverseproblem}
    \min_{\mathbf{w}} \Vert \phi(\tilde q;\mathbf{w}) - \mathbf{\bar n}\Vert
\end{equation}
for a suitable norm $\Vert \cdot  \Vert$.\footnote{In the literature, the search for such schemes is usually restricted to the space of weighted voting games. In principle, one could also consider the larger space of monotonic simple games. Neither is a vector space. This makes the inverse problem much harder than for general TU~games (see, e.g., \citeNP{Rojas/Sanchez:2016}).}

For large $m$ the SSI $\phi(v)$ of voting game $v=[\tilde q; w_1, \ldots, w_m]$ is often close to the relative weight vector $(w_1, \ldots, w_m)/\sum_i w_i$ (see, e.g., \citeNP{Jelnov/Tauman:2014}). Thus $\Vert \phi(\tilde q;\mathbf{\bar n}) - \mathbf{\bar n}\Vert$ tends to be small. So using population sizes as weights is a good practical default for implementing (\ref{eq:shock_case_solution}).

However, choosing $\mathbf{w}=\mathbf{\bar n}$ can involve considerable avoidable errors when $m$ is small, the distribution of constituency sizes is very skewed, or $q$ is close to 1. These cases are prone to pronounced non-proportionality of voting weight and voting power.
For instance, there exist only 9 structurally different weighted voting games  (up to isomorphisms) in case of $m=4$ and simple majority quota $q=0.5$. Numbers in the corresponding Shapley values $\phi(v)$ must be multiples of $1/4!=4.1\bar 6\%$ (cf.\ equation~(\ref{eq:Shapley})).
Exact proportionality to population shares of, say, 
$\mathbf{\bar n}=(42\%, 25\%, 24\%, 9\%)$ can, therefore, not be achieved~--
one must live with pivot probabilities which approximate $\mathbf{\bar n}$.
Default weights $(w_1, \ldots, w_4)=
\mathbf{\bar n}$ in this example induce approximate pivot probabilities of $\phi(v)=(50\%, 16.\bar 6\%, 16.\bar 6\%, 16.\bar 6\%)$.
This is arguably not a very satisfactory approximation.
In particular, it is more distant from $\mathbf{\bar n}$ than $\phi(v')=(41.\bar 6\%, 25\%, 25\%, 8.\bar 3\%)$, which would be induced by $(w'_1, \ldots, w'_4)=(40\%, 25\%, 25\%, 10\%)$.

Weights and power sometimes cannot be aligned to population figures even for large numbers of constituencies. In response to \citeN{Riker/Shapley:1968}, Robert \citeN[p.~221]{Nozick:1968} pointed to the following example with $q=0.5$: let an assembly consist of any odd number of legislators representing groups of equal size, and one legislator who represents a smaller group. Then each of the odd number of legislators must be given the same number of votes.
If the single legislator is given that weight, too, he or she would have power in excess of the size of the group; if given fewer weight, he or she would have no power at all.

Unfortunately, no useful bounds on the unavoidable gap to a given SSI target vector are known. Simple hill-climbing algorithms 
often deliver excellent results and good heuristic solutions exist if the relative quota $q$ is a variable rather than given (see \citeNP{Kurz/Napel:2014}). Still, one cannot rule out that these identify only a local minimum of the distance between the desired and the induced power vector.
For $m<9$, complete enumeration of voting games is the best option. \citeN{Kurz:2012} shows how \emph{integer linear programming} techniques can alternatively be brought to bear, but exact solutions are computationally demanding for $m>10$.
Even exact solutions to problem (\ref{eq:inverseproblem}) may involve non-negligible distances:
for instance, the Shapley vector with minimal $\Vert\cdot\Vert_1$-distance to $\mathbf{\bar n}=(49\%, 33\%, 9\%, 9\%)$ is $(41.\bar 6\%, 25\%, 25\%, 8.\bar 3\%)$.


\begin{figure}[p]
\centering
\begin{tabular}{ll}{\small (a)} \\ &
\includegraphics[width=0.825\textwidth]{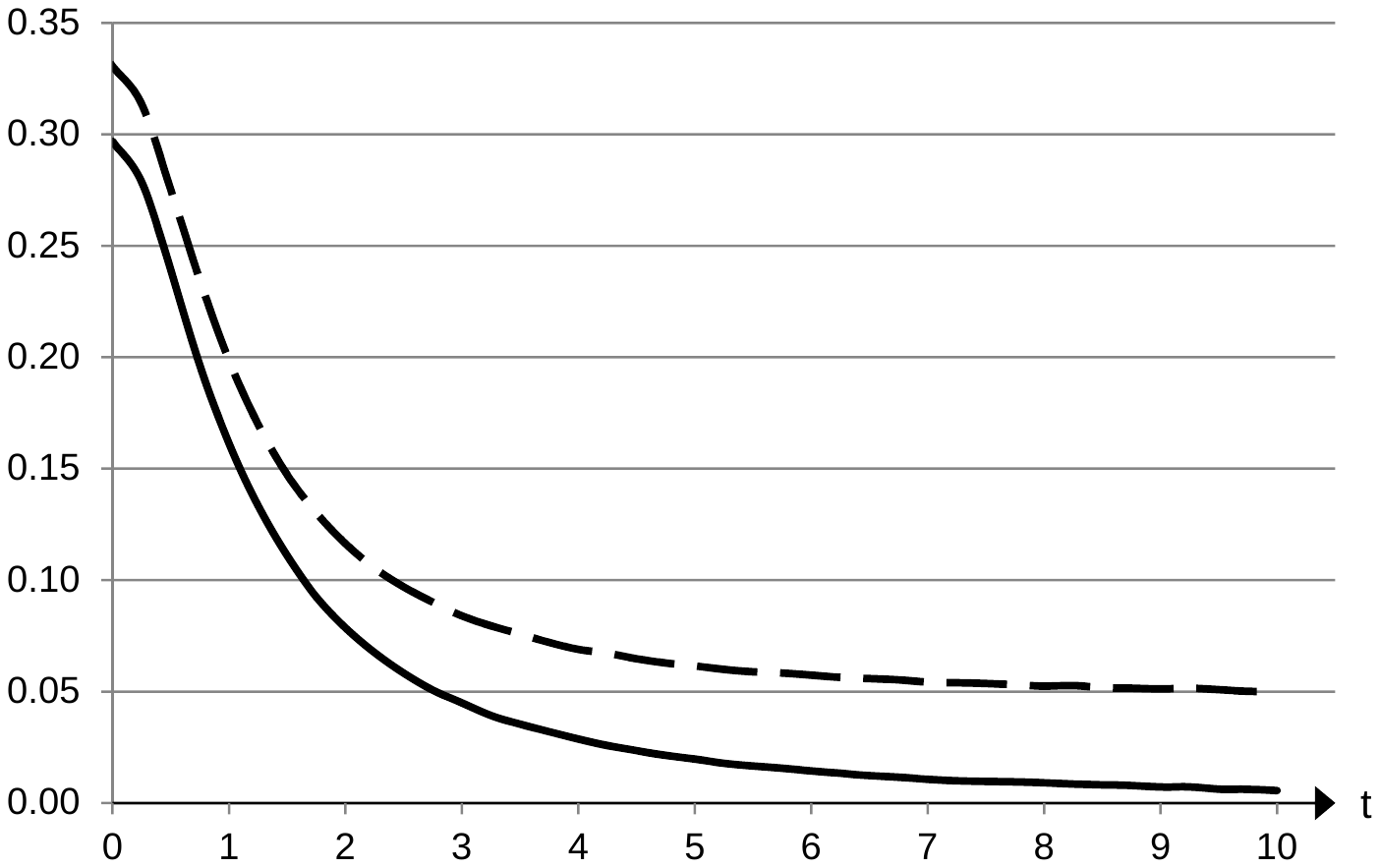}
\\
{\small (b)} \\ & \includegraphics[width=0.825\textwidth]{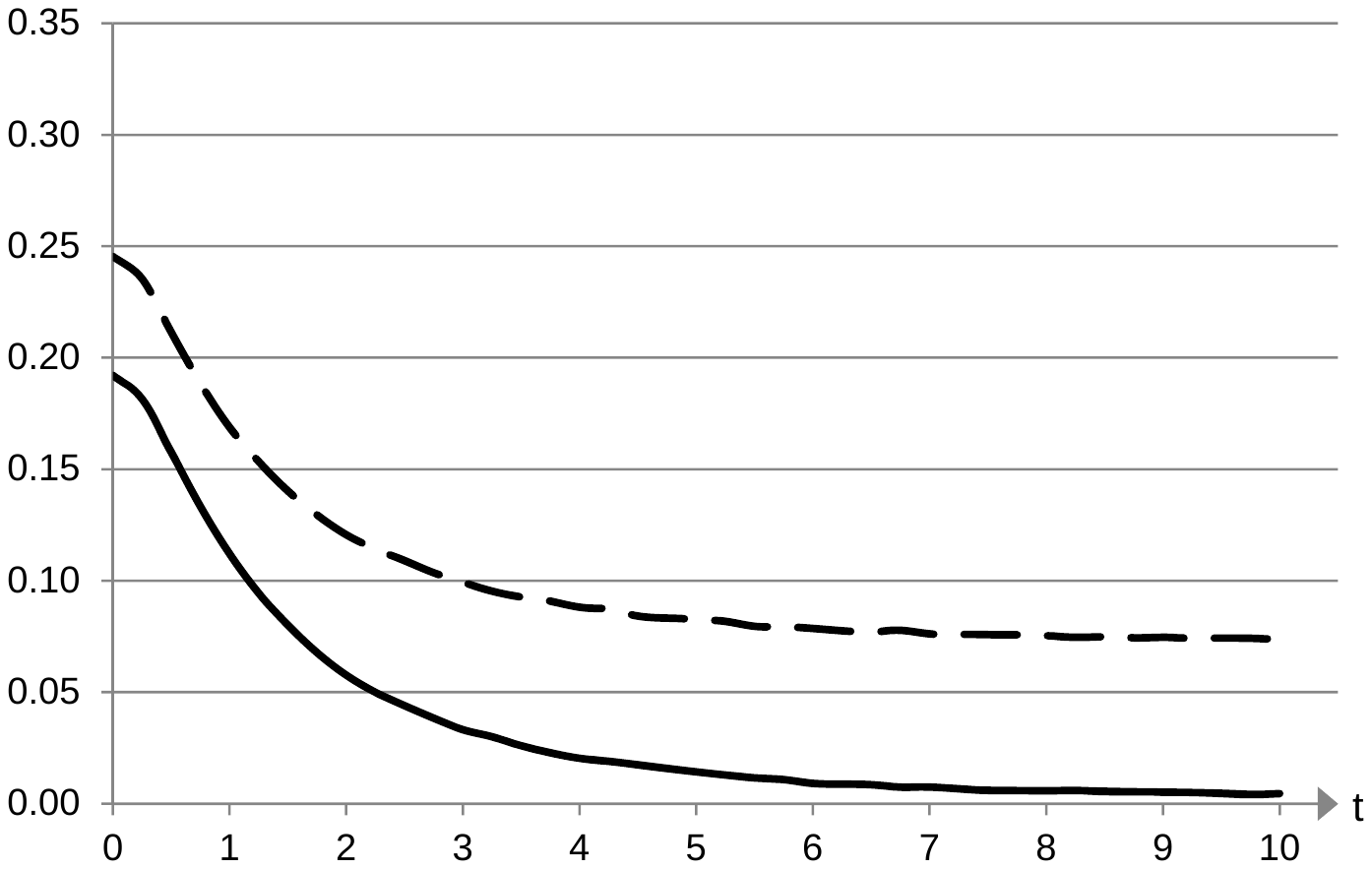}
\end{tabular}
  \caption{\small
Absolute deviation from perfectly fair representation for directly proportional (dashed line) and Shapley value-based weight allocations (solid line) with $n_1, \ldots, n_{28}$ defined by EU28 population data. Panel (a) and (b) consider $q = 0.5$ and $q = 0.74$, respectively.
}
  \label{fig:ComparisonEU28}
\end{figure}

Figure~\ref{fig:ComparisonEU28} illustrates that the asymptotic statement for $t\to \infty$ in Proposition~\ref{thm:main_result_for_shock_case} and hence Corollary~\ref{cor:equal_representation_shock_case} are already of use for small levels of $t$. The figure is based on weighted voting in the Council of the current EU with 28 member states under (a) a hypothetical quota of $q = 0.5$ and (b) the quota $q = 0.74$ (260 votes out of a total of 352) which was specified in the Treaty of Nice.\footnote{The treaty defined voting weights and a quota, and stipulated two other but essentially negligible criteria. The Nice rules can still be invoked in the EU until March 2017, when they will be replaced for good by the new voting system agreed in the Treaty of Lisbon.}
The lines respectively depict the $\Vert \cdot \Vert_1$-distance between individual influences and the perfectly fair democratic ideal of $(1/n, \ldots, 1/n)\in \realnumbers^n$ for weights following the linear Shapley rule (solid line) and the simple heuristic of choosing $\mathbf{w} = \mathbf{n}$ (dashed line).
\footnote{We considered $\epsilon^l\sim\mathbf{U}[-0.5, 0.5]$ and $\mu_i\sim\mathbf{N}(0,\sigma_H^2)$ with $\sigma_H^2 = 10^{-8}$. Estimates of the induced pivot probabilities $\pi_i(t)$ and hence deviations from the democratic ideal were obtained by Monte Carlo simulation. We used 
the Nelder-Mead method in order to solve the underlying inverse problems.}

The linear Shapley rule clearly outperforms simple population weights at any level of preference polarization.\footnote{In view of the limit results by \citeN{Neyman:1982} and \citeN{Lindner/Machover:2004}, it is noteworthy that there is still a noticeable advantage even for the relatively big number of 28 constituencies. The advantage can be expected to be higher for examples with smaller $m$.} 
The gap between representation according to the linear Shapley rule and perfectly fair representation narrows quickly as $t$ increases; it is already close to zero for $t\approx 10$.
One can also see that the lead of the Shapley-based weights over simple population weights is more pronounced for the higher vote threshold in panel~(b), in line with our earlier comments on the inverse problem.

\section{Concluding Remarks}\label{sec:Conclusion}
When Lloyd S.\ Shapley and his collaborators contemplated the problem of fair representation, they already mentioned proportionality to the Shapley value as a possible benchmark. \citeN{Riker/Shapley:1968} did not give it much emphasis, however, compared to a square root recommendation in the tradition of Penrose \citeyear{Penrose:1946,Penrose:1952} and \citeN{Banzhaf:1965}. The key reason to us seems their focus on perfectly exchangeable voters. 

This may actually be the most appropriate assumption in applications to, say, a federal state with high geographic mobility, like the US.
However, when constituencies correspond to entire nations as in case of the Council of the EU or the ECB~Governing Council, voters in a given constituency tend to share more historical experience, traditions, language, communication, etc.\ within constituencies than across (see \citeNP{Alesina/Spolaore:2003}).
Many lower key institutions with a delegate structure such as university senates, councils of NGOs, boards of sport clubs, etc.\ involve constituencies (faculties, divisions, and so on) whose   composition involves sorting  of like-minded individuals. 
Some preference similarity within and dissimilarity across constituencies thus often seems more plausible.

Our continuous rather than binary voting model then implies that equal expected influence on outcomes requires proportionality between a constituency's size and the respective probability~-- approximated by the Shapley value~-- of getting its way, i.e., of seeing its median voter's preferences implemented. We provided motivation for such proportionality by considering an individual's probability to be pivotal in his or her constituency, noting that it is the inverse of the respective population size.

This is \emph{not} the only possible motivation for proportional pivotality at the top tier.
As \shortciteN{Kurz/Maaser/Napel:2016} explain in more detail, one can also operationalize the influence of a given individual by considering the expected effect of \emph{participation}. Namely, every local voter almost always has influence on the position of the respective constituency median, and hence delegate, because abstention and consequent deletion from the considered sample would shift the realized median position. For instance, if a voter with an ideal point to the left of $\l_i$ is removed from the preference sample in $\mathcal{C}_i$, its median shifts to the right; a faithful delegate will then pursue a position $\l_i'>\l_i$ in the assembly. The expected size $|\l_i'-\l_i|$ of such a shift~-- and also a voter's incentives to turn out~-- falls in  $\mathcal{C}_i$'s population size and, specifically, can be shown to be asymptotically proportional to $1/n_i$. Therefore the same condition~(\ref{eq:OPOV_condition_restated}) follows.
Links from a constituency's pivotality to electoral campaign efforts and pork barrel funds also allow to arrive at Corollary~\ref{cor:equal_representation_shock_case} on equal treatment grounds.

In contrast to the square root result derived by \shortciteN{Kurz/Maaser/Napel:2016} for $t=0$, Corollary~\ref{cor:equal_representation_shock_case}  applies to arbitrary vote thresholds in the assembly. This admittedly involves a weaker notion in which the representative identified by equation~(\ref{eq:P_defined}) is `decisive' when $q>0.5$ compared to $q=0.5$. Still, it gives the linear Shapley rule additional robustness, which is appealing in view of the widespread use of supermajority rules in real decision making bodies.

At a normative level, the discrepancy between the findings for i.i.d.\ voters 
and positively affiliated voters 
raises a non-trivial question of practical philosophy: Which kinds of inter-constituency heterogeneity shall be acknowledged behind the `veil of ignorance'?
Constitutional design with a long-term perspective should arguably assume preferences to be distributed identically in all constituencies, even though historical patterns may suggest greater conservatism, religiosity, etc.\ for some constituencies rather than others.
There may analogously exist normative reasons outside the scope of our analysis for setting $t = 0$ even though $t>0$ is more plausible. 
Then, Riker and Shapley's \citeyear{Riker/Shapley:1968} main hunch about proportionality of the Shapley value to the square root of population sizes was right (at least for $q=0.5$). Otherwise, the linear rule which they discussed almost in passing provides the more 
``consistent criterion for `fair representation'{''}. 

\bigskip


\setlength{\labelsep}{-0.265cm}

\small

\newcommand{\noopsort}[1]{}

\end{document}